\documentclass{endm}
\usepackage{endmmacro}
\usepackage{graphicx}
\usepackage{microtype}

\newcommand{\heading}[1]{\smallskip\par\noindent{\bf #1}}

 \def\calR{{\cal R}} \def\calS{{\cal S}}

\def\cNP{\hbox{\rm \sffamily NP}}

\def\cXP{\hbox{\rm \sffamily XP}}
\def\cFPT{\hbox{\rm \sffamily FPT}}
\def\cAPX{\hbox{\rm \sffamily APX}}

\def\O{\mathcal{O}{}}

\def\PrColExt#1#2{{\textsc{PrColExt($#1,#2$)}}}

\def\int{\hbox{\rm \sffamily INT}}

\def\ca{\hbox{\rm \sffamily CARC}}

\def\chor{\hbox{\rm \sffamily CHOR}}

\def\split{\hbox{\rm \sffamily SPLIT}}

\newcommand{\graphs}[1]{\ensuremath{\hbox{\rm \sffamily $#1$-GRAPH}}}

\begin{document}

\begin{verbatim}\end{verbatim}\vspace{2.5cm}

\begin{frontmatter}

\title{Combinatorial Problems on $H$-graphs}

\author{Steven Chaplick}
\address{Lehrstuhl f\"ur Informatik~I, Universit\"at W\"urzburg, Germany,
 \\  Email: \texttt{first.last@uni-wuerzburg.de}}

\author{Peter Zeman}
\address{Department of Applied Mathematics, Faculty of Mathematics and Physics,\\ Charles
University in Prague, Czech Republic, \\ Email: \texttt{zeman@kam.mff.cuni.cz}} 

%

%

\begin{abstract}
Bir\'{o}, Hujter, and Tuza introduced the concept of $H$-graphs (1992),
intersection graphs of connected subgraphs of a subdivision of a graph $H$. They
naturally generalize many important classes of graphs, e.g., interval graphs and
circular-arc graphs. 
We continue the study of these graph classes by considering coloring, clique, and isomorphism problems on $H$-graphs. 

We show that for any fixed $H$ containing a certain 3-node, 6-edge multigraph as
a minor that the clique problem is \cAPX-hard on $H$-graphs and the isomorphism
problem is isomorphism-complete.  We also provide positive results on
$H$-graphs. Namely, when $H$ is a cactus the clique problem can be solved in
polynomial time. Also, when a graph $G$ has a Helly $H$-representation, the
clique problem can be solved in polynomial time.  Finally, we observe that one
can use treewidth techniques to show that both the $k$-clique and list
$k$-coloring problems are \cFPT\ on $H$-graphs. These \cFPT\ results apply more
generally to \emph{treewidth-bounded} graph classes where treewidth is bounded
by a function of the clique number.
\end{abstract}


\end{frontmatter}

\section{Introduction}

An intersection representation of a graph assigns a set to each vertex and uses
intersections of those sets to encode its edges. More formally, an intersection
representation $\calR$ of a graph $G$ is a collection of sets $\{R_v\}_{v \in
V(G)}$ such that $R_u \cap R_v \neq \emptyset$ if and only if $uv \in E(G)$.
Many important classes of graphs arise from restricting the sets $R_v$ to
geometric objects (e.g., intervals, convex sets).

We study \emph{$H$-graphs}, intersection graphs of connected subsets of a fixed
topological pattern given by a graph $H$, introduced by Bir\'{o}, Hujter, and
Tuza~\cite{biro1992precoloring}. We obtain new algorithmic results on 
clique, coloring, and isomorphism problem. In a companion paper~\cite{ChaplickTVZ16}, 
we studied recognition and dominating set problems on $H$-graphs. 
We begin with related graph classes.

\begin{figure}[b]
\centering
\includegraphics[scale=0.8]{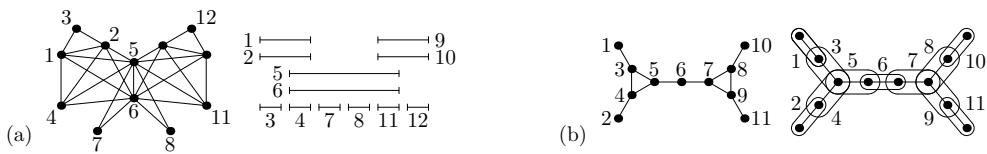}
\caption{(a) An interval graph and one of its interval representation. (b) A chordal
graph and one of its representation as an intersection graph of subtrees of a tree.}
\label{fig:int_chor_ex}
\end{figure}

\emph{Interval graphs} ($\int$) form one of the most studied and well-understood
classes of intersection graphs. In an \emph{interval representation}, each set
$R_v$ is a closed interval of the real line; see Fig.~\ref{fig:int_chor_ex}a.  A
primary motivation for studying interval graphs (and related classes) is the
fact that many important computational problems can be solved in linear time on
them; see for example~\cite{PQ_trees,chang1998efficient,lueker1979linear}.

A graph is \emph{chordal} when it does not have an induced cycle of length at
least four. Equivalently, as shown by Gavril~\cite{gavril1974intersection}, 
a graph is chordal if
and only if it can be represented as an intersection graph of subtrees of some
tree; see Fig.~\ref{fig:int_chor_ex}b. This immediately implies that
$\int$ is a subclass of the chordal graphs ($\chor$).
Some important problems (e.g., 
dominating set~\cite{booth1982dominating} and graph isomorphism~\cite{lueker1979linear}) 
are harder on chordal graphs than on interval graphs.



The \emph{split graphs} (\split) form an important subclass of chordal graphs. 
These are the graphs that can be partitioned into a clique and an independent set. Note
that every split graph is an intersection graph of subtrees
of a \emph{star} $S_d$, where $S_d$ is the complete bipartite graph $K_{1,d}$.

\emph{Circular-arc graphs} ($\ca$) generalize interval graphs by having 
each set $R_v$ be an arc of a circle.
A graph $G$ is a \emph{Helly circular-arc graph} if the
collection of circular arcs $\calR = \{R_v\}_{v \in V(G)}$ satisfies \emph{Helly
property}, i.e., in each sub-collection of $\calR$ whose sets pairwise
intersect, the common intersection is non-empty.  Interestingly, 
the coloring problem is \cNP-hard on Helly $\ca$ %
\cite{HellyCircArcColoring-1996}. 

\heading{$\mathbf{H}$-graphs.} 
Bir\'{o}, Hujter, and Tuza~\cite{biro1992precoloring} introduced
\emph{$H$-graphs}. Let $H$ be a fixed graph. A graph $G$ is an
\emph{intersection graph of $H$} if it is an intersection graph of connected
subgraphs of $H$, i.e., the assigned subgraphs $H_v$ and $H_u$ of $H$ share a
vertex if and only if $uv \in E(G)$.
A \emph{subdivision} $H'$ of a graph $H$ is obtained when the edges of $H$ are
replaced by internally disjoint paths of arbitrary lengths. A graph $G$ is a
\emph{topological intersection graph} of $H$ if $G$ is an intersection graph of
a subdivision $H'$ of $H$. We say that $G$ is an \emph{$H$-graph} and the
collection $\{H'_v : v \in V(G)\}$ of connected subgraphs of $H'$ is an
\emph{$H$-representation} of $G$. The class of all $H$-graphs is denoted by
$\graphs{H}$. We have the following relations: $\int =
\graphs{K_2}$, $\split \subsetneq \bigcup_{d = 2}^{\infty}\graphs{S_d}$, $\ca =
\graphs{K_3}$, and $\chor = \bigcup_{\{\rm Tree\}\ T}\graphs{T}$.
Moreover, for any pair of (multi-)graphs $H_1$ and $H_2$, if $H_1$ is a
\emph{minor} of $H_2$, then $\graphs{H_1} \subseteq \graphs{H_2}$. 
If $H_1$ is a subdivision of $H_2$, then $\graphs{H_1} = \graphs{H_2}$.

$H$-graphs were introduced in the context of the \emph{$(p,k)$ pre-coloring
extension problem}. Here, one is given a graph $G$ together
with a $p$-coloring of $W \subseteq V(G)$, and the goal is to find a proper
$k$-coloring of $G$ extending this \emph{pre-coloring}.  
\PrColExt{k}{k} has an $\cXP$ algorithm (in $k$ and $\|H\|$) for $\graphs{H}$.

%
%
	

\heading{Coloring $H$-graphs.}
From the above discussion, we note a dichotomy regarding computing a minimum coloring on \graphs{H}. 
Namely, if $H$ contains a cycle, then computing a minimum
coloring in $\graphs{H}$ is already \cNP-hard even for Helly $\graphs{H}$.  Additionally, when $H$
is acyclic, a minimum coloring can be computed in linear time since $\graphs{H}$ is a subclass of $\chor$.

\heading{Our Results.}
We prove that for any fixed $H$ containing a \emph{double triangle} (depicted in
Fig.~\ref{fig:comp2sub}) as a minor, the clique problem is \cAPX-hard on $H$-graphs and the
isomorphism problem is isomorphism-complete (see Section~\ref{sec:clique_and_iso_hardness}).  We
also provide positive results on $H$-graphs in Sections~\ref{sec:clique}~and~\ref{sec:treewidth}.
Namely, when a graph $G$ has a Helly $H$-representation, the clique problem can be solved in
polynomial time (see Theorem~\ref{thm:helly-h-clique}). 
Also, when $H$ is a cactus the clique problem can be solved in polynomial time (see Theorem~\ref{thm:clique-cactus}). 
Finally, we use treewidth techniques to show that both the $k$-clique and list $k$-coloring problems 
are \cFPT\ on $H$-graphs (see Propositions~\ref{pro:fpt-k-list-col}~and~\ref{pro:fpt-k-clique} respectively). 
These \cFPT\ results extend to \emph{treewidth-bounded} graph classes. 

%

%
%

\section{Finding Cliques in H-graphs}
\label{sec:clique}

This section concerns cases where the clique problem can be solved efficiently on 
\graphs{H}, for a fixed graph $H$.  First, we consider a case where we have
a ``nice'' representation but $H$ is arbitrary. Second, we restrict $H$
to be a cactus. 

\heading{Helly H-graphs.}
A \emph{Helly} $H$-graph $G$ has an $H$-representation 
$\{H'_v : v \in V(G)\}$ such that the collection $\calS =
\{V(H'_v) : v \in V(G)\}$ satisfies the \emph{Helly property}, i.e., for each
sub-collection of $\mathcal{S}$ whose sets pairwise intersect, their common
intersection is non-empty. 
Notice that, when $H$ is a tree, every $H$-representation satisfies the Helly
property.  Furthermore, when a graph $G$ has a Helly $H$-representation, 
we obtain the following relationship between the size of $H$ and the number
of maximal cliques in $G$. 

\begin{lemma}\label{lem:helly-h-cliques}
Each Helly $H$-graph $G$ has at most $|V(H)| + |E(H)|\cdot|V(G)|$ maximal cliques.
\end{lemma}
\begin{proof}
Let $H'$ be a subdivision of $H$ such that $G$ has a Helly $H$-representation
$\{H'_v : v \in V(G)\}$. Note that, for each maximal clique $C$ of $G$,
$\bigcap_{v \in C} H'_v \neq \emptyset$, i.e., $C$ corresponds to a node $x_C$
of $H'$. For every edge $xy \in E(H)$, we consider the corresponding path $P =
(x,x_1,\dots,x_k,y)$ in $H'$. Let $G_{xy}$ be the subgraph of $G$ formed by  
maximal cliques of $G$ which ``occur'' on $P$. The graph $G_{xy}$ is a 
Helly cicular-arc graph. Now, since Helly circular arc graphs have at most linearly 
many maximal cliques~\cite{gavril1974algorithms}, $G$ has at most 
$|V(H)| + |E(H)| \cdot |V(G)|$ maximal cliques.
\end{proof}

We can now use Lemma~\ref{lem:helly-h-cliques}
to find the largest clique in $G$
in polynomial time. In fact, we can do this without needing to compute a
representation of $G$. In particular, the maximal cliques of a graph can be
enumerated with polynomial delay~\cite{makino2004new}. Thus, since $G$ has at
most linearly many maximal cliques, we can simply list them all in polynomial
time and report the largest, i.e., if the enumeration process produces too many
maximal cliques, we know that $G$ has no Helly $H$-representation. This provides
the following theorem. 

\begin{theorem}
\label{thm:helly-h-clique}
The clique problem is polytime solvable on Helly $H$-graphs. 
\end{theorem}

Note that some co-bipartite circular arc graphs have have exponentially many 
maximal cliques 
and as such are not contained in Helly $H$-graphs for any fixed $H$. 
However, the clique problem is polytime solvable on $\ca$
~\cite{Hsu85-max-clique}. 

\heading{Cactus-graphs.}
The clique problem is efficiently solvable on chordal graphs~\cite{agt} and
circular arc graphs~\cite{Hsu85-max-clique}. In particular, when $H$
is either a tree or a cycle, the clique problem can be solved in polynomial
time independent of the size of $H$. 
In Theorem~\ref{thm:clique-cactus}, we observe that these results easily generalize 
to the case when $G$ is in \graphs{H} for some cactus $H$.  
With this in mind, we say that such a graph $G$ belongs to the class \graphs{cactus}, 
where $\graphs{cactus} = \bigcup \{\graphs{H} : H \textnormal{ is a cactus.}\}$. 

To prove the result we will use the \emph{clique-cutset decomposition} -- which is
defined as follows.  
A \emph{clique-cutset} of a graph $G$ is a clique $K$ in $G$ such that $G \setminus K$ has more connected components than $G$. An \emph{atom} is a graph without a clique-cutset.
An \emph{atom of a graph $G$} is an induced subgraph $A$ of $G$ which is an atom. A 
\emph{clique-cutset decomposition} of $G$ is a set $\{A_1, \ldots, A_k\}$ of atoms of $G$
such that $G = \bigcup_{i=1}^k A_i$ and for every $i,j$, $V(A_i) \cap V(A_j)$ is either empty or 
induces a clique in $G$. 
Algorithmic aspects of clique-cutset decompositions were studied by 
Whitesides~\cite{Whi1984} and Tarjan~\cite{Tar1985}. In particular, if $k\leq n$, then for any graph $G$ a 
clique-cutset decomposition $\{A_1, \ldots, A_k\}$ of $G$ can be computed in 
$O(n\cdot(n+m))$~\cite{Tar1985}. Additionally, to solve 
the clique problem on a graph $G$ it suffices to solve it for each atom of $G$ from a 
clique-cutset decomposition~\cite{Whi1984,Tar1985}. 
Theorem~\ref{thm:clique-cactus} now follows from the following easy lemma and the fact that 
the clique problem can be solved in polynomial time on circular arc 
graphs~\cite{Hsu85-max-clique}. 

\begin{lemma}
\label{obs:cactus-atoms}
If $G \in \graphs{cactus}$, then each atom $A$ of $G$ is in $\ca$.
\end{lemma}
\begin{proof}
Consider an $H$-representation $\{H_v : v \in V(G)\}$ of $G$ where $H$ is a cactus. 
Now let $H|_A = \bigcup_{v \in V(A)} H_v$. Clearly, if $H|_A$ is a path or a cycle, then 
we are done. Otherwise, $H|_A$ must contain a cut-node $x$. 
Let $C_1, \ldots, C_t$ be the components of $H|_A \setminus \{x\}$, and let $S$ 
be the vertices of $A$ whose representations contain $x$. 
Note that $S$ is a clique in $A$. Moreover, since $A$ is an atom, $S$ is not a 
clique-cutset. 
Thus, there is a component $C_j$ such that the subgraph $H'$ of $H$ induced by 
$V(C_j) \cup \{x\}$ provides a representation of $A$. In particular, if $H'$ is
either a cycle or a path we are again done. Moreover, when $H'$ is neither a path nor a 
cycle, repeating this argument on $H'$ provides a smaller subgraph of $H$ on which $A$ 
can be represented, i.e., this eventually produces either a path or cycle.
\end{proof}

\begin{theorem}
\label{thm:clique-cactus}
The clique problem can be solved in polynomial time on the class \graphs{cactus}. 
\end{theorem}

\newcommand{\Subd}[1]{\ensuremath{\hbox{\rm \sffamily SUBD}_#1}}
\newcommand{\compSubd}[1]{\ensuremath{\overline{\Subd{#1}}}}

\section{Clique and Isomorphism Hardness Results} 
\label{sec:clique_and_iso_hardness}

To obtain our hardness results we show that there are graphs $H$ such that 
the complement of a $2$-subdivision of every graph is an $H$-graph. The 
$2$-subdivision of a graph $G$ is the result of subdividing every 
edge of $G$ twice. The complement of a graph $G$ is denoted by $\overline{G}$.
We use $\Subd{2}$ to denote the class of all 2-subdivisions of graphs
and $\compSubd{2}$ to denote their complements.

This seemingly esoteric family of graphs is interesting for two reasons. The
first is that graph isomorphism is closed under $k$-subdivision and complement
operations. Thus, isomorphism testing in $\compSubd{2}$ is as hard as it is for
general graphs, i.e., the class $\compSubd{2}$ is \emph{isomorphism-complete}.
The second is that the clique problem is \cAPX-hard on \compSubd{2}. More
specifically, Chleb\'ik and Chleb\'ikov\'a~\cite{ChlebikC07-subdiv-hardness}
proved that the maximum independent set problem is \cAPX-hard on the class of
$2k$-subdivisions of 3-regular graphs for any fixed integer $k \geq 0$; in
particular, for 2-subdivisions.  Thus, showing that $\compSubd{2} \subseteq
\graphs{H}$ for a fixed $H$, implies that the maximum clique problem is
$\cAPX$-hard on $\graphs{H}$ and that $\graphs{H}$ is isomorphism-complete. 


\begin{theorem}
\label{thm:double-triangle->Subd2_inside}
If $H$ contains the graph in Fig.~\ref{fig:comp2sub}a as a minor, then
$\compSubd{2} \subseteq \graphs{H}$.
\end{theorem}
\begin{proof}
Since $H$ contains the graph in Fig.~\ref{fig:comp2sub} as a minor, it can 
be partitioned into three connected subgraphs $H_1$, $H_2$, $H_3$ such that 
there are at least two edges connecting $H_i$ and $H_j$ for each $i\neq j$. 
For every graph $G$, we show that the complement of its $2$-subdivision has 
and $H$-representation.

The construction proceeds similarly to the constructions used by Francis et
al.~\cite{Francis2013-multiple-interval}, and we borrow their convenient
notation.  Let $G$ be a graph with vertex set $\{v_1, \ldots, v_n\}$ and edge
set $\{e_1, \ldots, e_m\}$. If $e_k \in E(G)$ and $e_k = v_iv_j$ where $i < j$,
we define $l(k) = i$ and $r(k) = j$ (as if $v_i$ and $v_j$ were respectively the
\emph{left} and \emph{right} ends of $e_k$). In the 2-subdivision $G^*$
of $G$, the edge $e_k$ of $G$ is replaced by the path $(v_{l(k)}, a_k,
b_k, v_{r(k)})$; see Fig.~\ref{fig:comp2sub}a and Fig.~\ref{fig:comp2sub}b. 

\begin{figure}[t]
\centering
\includegraphics[width=\linewidth]{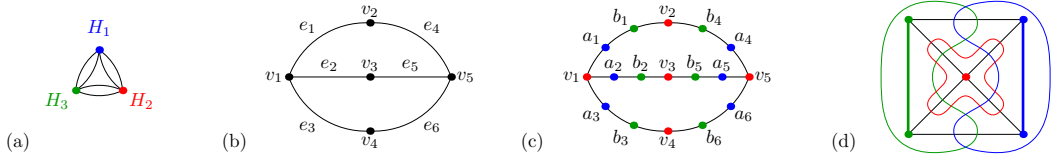}
\caption{(a) The \emph{double triangle} graph. (b) A graph $G$. (c) The
2-subdivision
$G^*$ of $G$. A three-clique cover of $\overline{G^*}$ is indicated by colors.
(d) The $4$-wheel graph (which contains the double triangle as a minor) and a \emph{sketch} 
of our $H$-representation of $\overline{G^*}$. For example, the edges between the green
clique and the blue clique are represented where the green and blue regions
intersect.}
\label{fig:comp2sub}
\end{figure}

Note that $\overline{G^*}$ can be covered by three cliques, i.e.,
$C_v = \{v_1, \ldots, v_n\}$, $C_a = \{a_1, \ldots, a_m\}$, and $C_b = \{b_1,
\ldots, b_m\}$. We now describe a subdivision $H'$ of $H$ which admits 
an $H$-representation $\{H'_v : v \in V(\overline{G^*})\}$ of
$\overline{G^*}$. We obtain $H'$ by subdividing the six  edges
connecting $H_1$, $H_2$, and $H_3$. Specifically: 
\begin{itemize}
\item we $n$-subdivide the two edges connecting $H_1$ to $H_2$ to obtain two paths $P_{12} = (\alpha_0, \alpha_1, \ldots,$ $\alpha_n, \alpha_{n+1})$, $Q_{12} = (\beta_0, \beta_1, \ldots, \beta_n, \beta_{n+1})$ where $\alpha_0, \beta_0 \in H_1$ and $\alpha_{n+1}, \beta_{n+1} \in H_2$, and
\item we $n$-subdivide the two edges connecting $H_1$ to $H_3$ to obtain two paths $P_{13} = (\gamma_0, \gamma_1, \ldots,$ $\gamma_n, \gamma_{n+1})$, $Q_{13} = (\eta_0, \eta_1, \ldots, \eta_n, \eta_{n+1})$ where $\gamma_0, \eta_0 \in H_1$ and $\gamma_{n+1}, \eta_{n+1} \in H_2$. 
\item $m$-subdivide the two edges connecting $H_2$ and $H_3$ to obtain two paths $P_{23} = (\mu_0, \mu_1, \ldots,$ $\mu_m, \mu_{m+1})$, $Q_{23} = (\nu_0, \nu_1, \ldots, \nu_m, \nu_{m+1})$ where $\mu_0, \nu_0 \in H_2$ and $\mu_{m+1}, \eta_{m+1} \in H_2$. 
\end{itemize} 

We now describe each $H_{v_i}$, $H_{a_j}$ and $H_{b_j}$. The idea is that
$H'_{v_i}$ will contain $H_1$ and extend from the ``start'' of $P_{12}$ up to
the position $i$, and from the ``start'' of $Q_{12}$ up to position $(n-i)$.
From the other side, each $H'_{a_j}$ will contain $H_2$ and extend from the
``end'' of $P_{12}$ down to position $(l(j)+1)$, and from the end of $Q_{12}$
down to position $(n-l(j)+1)$; an example is sketched in
Fig.~\ref{fig:comp2sub}d. In this way, we ensure that $H'_{a_j}$ does not
intersect $H'_{v_{l(j)}}$ while $H'_{a_j}$ does intersect every $H'_{v_i}$ for
$i \neq l(j)$. The other pairs proceed similarly, and we describe the subgraphs
$H_{v_i}, H_{a_j}, H_{b_j}$ for each $i \in \{1,\dots,n\}$ and $j \in
\{1,\dots,m\}$ as follows: 
\begin{itemize}
\item $H'_{v_i} = H_1 \cup \{\alpha_1, \ldots, \alpha_i\} \cup \{\beta_1, \ldots, \beta_{n-i}\} \cup \{\gamma_1, \ldots, \gamma_i\} \cup \{\eta_1, \ldots, \eta_{n-i}\}$.
\item $H'_{a_j} = H_2 \cup \{\alpha_n, \ldots, \alpha_{l(j)+1}\} \cup \{\beta_n, \ldots, \beta_{n-l(j)+1}\} \cup \{\mu_1, \ldots, \mu_j\} \cup \{\nu_1, \ldots, \nu_{m-j}\}$. 
\item $H'_{b_j} = H_3 \cup \{\gamma_n, \ldots, \gamma_{r(j)+1}\} \cup \{\eta_n, \ldots, \eta_{n-r(j)+1}\} \cup \{\mu_m, \ldots, \mu_{j+1}\} \cup \{\nu_m, \ldots, \nu_{m-j+1}\}$. 
\end{itemize} 
\end{proof}




Recall that, Theorem~\ref{thm:clique-cactus} states that the clique problem can
be solved in polynomial time on cactus-graphs. 
Thus, the open cases which remain are when $H$ is not a cactus (i.e., $H$
contains a diamond as a minor), but $H$ does not satisfy the conditions of
Theorem~\ref{thm:double-triangle->Subd2_inside}.  On the other hand, while the
isomorphism problem can be solved in linear time on interval graphs and Helly
circular-arc graphs~\cite{szwarcfiter2013isomorphism}, it is
isomorphism-complete on split graphs~\cite{lueker1979linear}.
Many questions remain open for the
complexity status of the isomorphism problem on \graphs{H}, even for the
simplest non-chordal case, circular-arc
graphs~\cite{szwarcfiter2013isomorphism}.


\newcommand{\tw}{tw}
\newcommand{\Oh}{\O}
\newcommand{\td}{TD}

\section{FPT Results via Treewidth-bounded Graph Classes}
\label{sec:treewidth}

In this section we discuss the concept of \emph{treewidth-bounded} graph classes. 
We will use the fact that the class $\graphs{H}$ has ``well-behaved'' treewidth (see
Lemma~\ref{lem:bounded_tw}) together with some observations about
more general \emph{treewidth-bounded} graph classes to study
optimization problems on $\graphs{H}$. 

\emph{Treewidth} was introduced by Robertson and
Seymour~\cite{robertson1984graph}. A {\em tree decomposition} of a graph
$G=(V,E)$ 
is a pair $({X}, T)$, where $T$ is a tree and
${X}=\{{X}_i \mid i\in V(T) \}$ is a family of subsets of $V$, called
\emph{bags}, such that 
(1) for all $v \in V$, the set of nodes $T_v = \{i \in V(T) \mid v \in {X}_i\}$ induces a non-empty connected subtree of $T$, and
(2) for each edge $e=\{u,v\} \in E(G)$ there exists $i \in V(T)$ such that both $u$ and $v$ are in ${X}_i$.
The maximum of
$|{X}_i|-1$, $i\in V(T)$, is called the {\em width} of the tree decomposition.
The {\em treewidth}, $\tw(G)$, of a graph $G$ 
is the minimum width
over all tree decompositions of $G$.

An easy lower bound on the treewidth
of a graph $G$ is the size of the largest clique in $G$, i.e., its clique number
$\omega(G)$. This follows from the fact that each edge of $G$ belongs to
 some bag of $T$ and that a collection of pairwise intersecting subtrees
of a tree must have a common intersection (i.e., they satisfy the Helly
property). With this in mind, we say that a graph class $\mathcal{G}$ is
\emph{treewidth-bounded} if there is a function $f: \mathbb{N}\rightarrow
\mathbb{N}$ such that for every $G \in \mathcal{G}$, $tw(G) \leq f(\omega(G))$.
This concept generalizes the idea of $\mathcal{G}$
being \emph{$\chi$-bounded}, namely, that the \emph{chromatic number} $\chi(G)$
of every graph $G \in \mathcal{G}$ is bounded by a function of the clique number
of $G$. In particular, the chromatic number of a graph $G$ is bounded by its
treewidth since a tree decomposition $G$ is a tree representation of a chordal
supergraph $G'$ of $G$ where $\omega(G') = \tw(G)+1$, i.e., $\chi(G') =
\tw(G)+1$ since chordal graphs are perfect. 
It was recently shown that the graphs which do not contain
\emph{even holes} (i.e., cycles of length $2k$ for any $k\geq 2$) and \emph{pans}
(i.e., cycles with a single pendent vertex attached) as induced subgraphs are 
treewidth bounded by $f(\omega) = \frac{3}{2}\omega -1$~\cite{cameron2015structure}. 

For a function $f: \mathbb{N} \rightarrow \mathbb{N}$, we use $\mathcal{G}_f$ to
denote the class of graphs $G$ where $tw(G) \leq f(\omega(G))$.  
Each class \graphs{H} is known to be a subclass of $\mathcal{G}_f$ for certain linear functions $f$, as in the following lemma of Biro et
al~\cite{biro1992precoloring}. 

\begin{lemma}\label{lem:bounded_tw}
\cite{biro1992precoloring}
For every $G \in \graphs{H}$, $\tw(G) \leq (\tw(H)+1) \cdot \omega(G) -1$, i.e.,
$\graphs{H} \subseteq \mathcal{G}_{f_H}$ where $f_H(\omega) = (\tw(H)+1) \cdot
\omega -1$. 
\end{lemma}

We now apply the existing literature to describe
the computational complexity of $k$-coloring problems as well as the $k$-clique
problem on treewidth-bounded graph classes, and, in particular, the $\graphs{H}$
classes. 



For each fixed $k \geq 3$, it is also known that testing for a 
\emph{$(k,k)$-pre-colouring extension} in the class \graphs{H} can be done in \cXP~time \cite{biro1992precoloring}. 
They use Lemma~\ref{lem:bounded_tw} together with a simple argument to obtain
their result.  We use a similar argument together with a more recent result 
regarding bounded treewidth graphs to observe that an even
more general problem, \emph{list $k$-coloring} (where each list is a subset of $\{1, \ldots, k\}$), 
is \cFPT\ on any
treewidth-bounded graph class, and as such also on \graphs{H}, i.e., 
Proposition~\ref{pro:fpt-k-list-col}. We first show that the
$k$-clique problem is \cFPT\ on any treewidth-bounded graph class. 


\begin{proposition}\label{pro:fpt-k-clique}
For any computable function $f: \mathbb{N} \rightarrow \mathbb{N}$, the
$k$-clique problem can be solved in $\Oh((5\cdot f(k))^{5\cdot f(k)}\cdot n)$ time on
$\mathcal{G}_f$. Thus, for $\graphs{H}$, the $k$-clique problem can
be solved in $\Oh((5\cdot\tw(H)\cdot k)^{5\cdot\tw(H)\cdot k}\cdot n)$ time.   
\end{proposition}
\begin{proof}
To test if $G$ contains a $k$-clique, we first try to generate a tree
decomposition of $G$ with width roughly $f(k)$ via a recent algorithm~\cite{DBLP:journals/siamcomp/BodlaenderDDFLP16} which, for any given graph $G$ and number $t$, provides a tree decomposition of width at most $5\cdot t$ or states that the treewidth of $G$ is larger than $t$ -- this algorithm runs in $2^{\Oh(t)}\cdot n$ time. If this algorithm provides tree decomposition, we use it to test whether $G$ has a $k$-clique in $\Oh((5\cdot f(k))^{5 \cdot f(k)} \cdot n)$ time via a known algorithm~\cite{courcelle1992monadic}. 
If not, then $G$ must contain a $k$-clique, and we are done. 
\end{proof}

\begin{proposition}\label{pro:fpt-k-list-col}
For any function $f: \mathbb{N} \rightarrow \mathbb{N}$, the list-$k$-coloring
problem can be solved in $\Oh(((5\cdot f(k))^{5\cdot f(k)}+k^{5\cdot f(k)+2})\cdot n)$ time on
$\mathcal{G}_f$. Thus, for $\graphs{H}$, the list-$k$-coloring problem
can be solved in $\Oh(((5\cdot \tw(H)\cdot k)^{5\cdot \tw(H)\cdot k}+k^{(5\cdot \tw(H)\cdot k)+2})\cdot n)$ time.   
\end{proposition}
\begin{proof}
For fixed $k$, clearly, if $G$ contains a clique of size $k+1$ then $G$ has no
$k$-coloring, i.e., no list-$k$-COL regardless of the lists. 
We use Proposition~\ref{pro:fpt-k-clique} to test for such a clique, and reject
if one is found. 
Otherwise, we have a $5\cdot f(k)$ width tree decomposition, and this time use it to
solve the list-$k$-COL problem via the known $\Oh(n\cdot k^{t+2})$ time
algorithm when given a width $t$ tree decomposition~\cite{JansenScheffler-k-list-col-1997}, i.e.,
list-$k$-COL can be solved in $\Oh(n \cdot k^{5\cdot k\cdot\tw(H) +2})$-time on
\graphs{H}. 
\end{proof}
\bibliographystyle{endm}
\bibliography{h-graphs_opt}

\end{document}